\documentclass{article}
\usepackage{spconf,amsmath,graphicx,hyperref}
\usepackage{flushend}

\hypersetup{
    colorlinks=true,
    allcolors=black,
}

\usepackage[acronym,shortcuts]{glossaries}
\newacronym[firstplural=spatio-temporal covariance matrices (STCMs)]{STCM}{STCM}{spatio-temporal covariance matrix}
\newacronym{BLCMP}{BLCMP}{binaural linearly constrained minimum power}
\newacronym{RIR}{RIR}{room impulse response}
\newacronym{LCMP}{LCMP}{linearly constrained minimum power}
\newacronym{wBLCMP}{wBLCMP}{weighted binaural linearly constrained minimum power}
\newacronym{STFT}{STFT}{short-time Fourier transform}
\newacronym{CTF}{CTF}{convolutive transfer function}
\newacronym{MTF}{MTF}{multiplicative transfer function}
\newacronym{RTF}{RTF}{relative transfer function}
\newacronym{ATF}{ATF}{acoustic transfer function}
\newacronym{MCLP}{MCLP}{multi channel linear prediction}
\newacronym{MPDR}{MPDR}{minimum power distortionless response}
\newacronym{MVDR}{MVDR}{minimum variance distortionless response}
\newacronym{LCMV}{LCMV}{linearly constrained minimum variance}
\newacronym{WPD}{WPD}{weighted power minimization distortionless response}
\newacronym{WPE}{WPE}{weighted prediction error}
\newacronym{TVG}{TVG}{time-varying complex circular Gaussian}
\newacronym{MISO}{MISO}{multiple-input single-output}
\newacronym{MIMO}{MIMO}{multiple-input multiple-output}
\newacronym{SPP}{SPP}{speech presence probability}
\newacronym{PESQ}{PESQ}{perceptual evaluation of speech quality}
\newacronym{FWSSNR}{FWSSNR}{frequency-weighted segmental signal-to-noise ratio}
\newacronym{CW}{CW}{covariance whitening}
\newacronym{CS}{CS}{covariance subtraction}
\newacronym{VAD}{VAD}{voice activity detection}
\newacronym{SAD}{SAD}{source activity detection}
\newacronym{UCB}{UCB}{unified convolutional beamformer}
\newacronym{IRLS}{IRLS}{iteratively reweighted least squares}
\newacronym{MFMVDR}{MFMVDR}{multi-frame minimum variance distortionless response}
\newacronym{BMFMVDR}{BMFMVDR}{binaural MFMVDR}
\newacronym{TCN}{TCN}{temporal convolutional network}
\newacronym{DNN}{DNN}{deep neural network}
\newacronym{GRU}{GRU}{gated recurrent unit}
\newacronym{RNN}{RNN}{recurrent neural network}
\newacronym{SNR}{SNR}{signal-to-noise ratio}
\newacronym{DRC}{DRC}{dynamic range compression}
\newacronym{HGR}{HGR}{half-gain rule}
\newacronym{CEC1}{CEC1}{Clarity Enhancement Challenge}
\newacronym{MBDRC}{MBDRC}{multi-band dynamic range compressor}
\newacronym{MBSTOI}{MBSTOI}{modified binaural short-term objective intelligibility}
\newacronym{SD-SDR}{SD-SDR}{scale-dependent signal-to-distortion ratio}
\newacronym{IFC}{IFC}{inter-frame correlation}
\newacronym{BTE}{BTE}{behind-the-ear}
\newacronym{SIR}{SIR}{signal-to-interferer ratio}
\newacronym{SINR}{SINR}{signal-to-interferer-and-noise ratio}
\newacronym{SRR}{SRR}{signal-to-reverberation ratio}
\newacronym{DUR}{DUR}{desired-to-undesired ratio}
\newacronym{wMPDR}{wMPDR}{weighted \gls{MPDR}}
\newacronym{SVD}{SVD}{singular value decomposition}
\newacronym{LS}{LS}{least-squares}
\newacronym{CBW}{CBW}{covariance blocking and whitening}
\newacronym{SRI}{SRI}{successive \gls{RTF} identification}
\newacronym{BOP-SRI}{BOP-SRI}{blind oblique projection \gls{SRI}}
\newacronym{BOP}{BOP}{blind oblique projection}
\newacronym{EBOP}{EBOP}{extended blind oblique projection}
\newacronym{CWu}{CWu}{\gls{CW} with the undesired covariance matrix}
\newacronym[\glslongpluralkey=power spectral densities]{PSD}{PSD}{power spectral density}
\newacronym{HA}{HA}{Hermitian angle}
\newacronym{MHA}{MHA}{mean Hermitian angle}
\newacronym{WMHA}{WMHA}{weighted mean Hermitian angle}
\newacronym{GMSC}{GMSC}{generalized magnitude-squared coherence}
\newacronym{LTASS}{LTASS}{long-term average speech spectrum}
\newacronym{EM}{EM}{expectation maximization}
\newacronym{CFA}{CFA}{confirmatory factor analysis}
\newacronym{DOA}{DOA}{direction-of-arrival}
\newacronym{BOPO}{BOPO}{BOP using orthogonal additional vectors}
\newacronym{BOP-S}{BOP-S}{BOP with noise subtraction}
\newacronym{BOP-W}{BOP-W}{BOP with noise whitening}
\newacronym{BOPO-S}{BOPO-S}{BOPO with noise subtraction}
\newacronym{BOPO-W}{BOPO-W}{BOPO with noise whitening}
\newacronym{MAE}{MAE}{mean absolute error}
\usepackage{amsmath,amssymb,amsfonts,bm,nicefrac,siunitx}
\usepackage{cite}
\usepackage{cleveref}
\crefname{equation}{}{}
\usepackage{tikz}
\usetikzlibrary{shapes.geometric, arrows}
\usepackage{booktabs}
\usepackage{enumitem}

\newlength{\abovesectionvspace}
\newlength{\belowsectionvspace}
\newlength{\abovesubsectionvspace}
\newlength{\belowsubsectionvspace}
\newlength{\figcapvspace}
\newlength{\tablecapvspace}

\usepackage[subtle]{savetrees}

\title{DNN-Based Online Source Counting Based on \\ Spatial Generalized Magnitude Squared Coherence}

\name{Henri Gode, Simon Doclo\thanks{
  This work was funded by the Deutsche Forschungsgemeinschaft (DFG, German Research Foundation)
  under Germany's Excellence strategy -- Project ID 390895286 -- EXC 2177/1.
  }}
\address{
  Department of Medical Physics and Acoustics and Cluster of Excellence Hearing4all,
  \\ Carl von Ossietzky Universität Oldenburg, Germany,
  e-mail: \href{mailto:henri.gode@uni-oldenburg.de}{\texttt{henri.gode@uni-oldenburg.de}}%
}

\begin{document}
\ninept
\maketitle
%
% The column width is: \the\columnwidth
% The line width is: \the\linewidth
% The text width is: \the\textwidth

\begin{abstract}
  The number of active sound sources is a key parameter in many acoustic signal processing tasks,
  such as source localization, source separation, and multi-microphone speech enhancement.
  This paper proposes a novel method for online source counting
  by detecting changes in the number of active sources based on spatial coherence.
  The proposed method exploits the fact that a single coherent source in spatially white background noise yields high spatial coherence,
  whereas only noise results in low spatial coherence.
  By applying a spatial whitening operation,
  the source counting problem is reformulated as a change detection task,
  aiming to identify the time frames when the number of active sources changes.
  The method leverages the \acrlong{GMSC} as a measure to quantify spatial coherence,
  providing features for a compact neural network trained to detect source count changes framewise.
  Simulation results with binaural hearing aids in reverberant acoustic scenes
  with up to $4$ speakers and background noise
  demonstrate the effectiveness of the proposed method for online source counting.
\end{abstract}

\glsresetall

\vspace{\abovesectionvspace}
\section{Introduction}
\label{sec:intro}
\vspace{\belowsectionvspace}

Enhancing speech quality and intelligibility
in complex acoustic environments with multiple competing sound sources
is a primary goal of assistive listening devices, such as binaural hearing aids.
To achieve this, advanced multi-microphone algorithms
are employed for tasks such as source localization,
source separation,
and speech enhancement~\cite{doclo_multichannel_2015,gannot_consolidated_2017,pertila2018multichannel,grumiaux2022survey}.
A fundamental prerequisite for many of these algorithms
is knowledge about the number of active sound sources.
In dynamic scenarios where sources may activate and deactivate over time,
it is crucial to perform online source counting,
i.e., relying only on past information.

Several methods have been proposed for counting active sound sources.
Single-microphone methods typically rely on \glspl{DNN} to estimate the source count
directly from the mixture~\cite{kinoshita2018listening, stoter2018countnet,Chetupalli_speakercount_2023}.
In contrast, multi-microphone methods are able to leverage spatial information.
Some methods perform clustering on spatial features such as \acrlong{DOA} estimates,
often requiring a-priori knowledge of the microphone array geometry
and hence limiting their applicability~\cite{pavlidi2013real,wang2016iterative,azcarreta2018permutation,hafezi2019spatial}.
Geometry-agnostic methods have been proposed based on the eigenvalue distribution of spatial correlation
or coherence matrices~\cite{Laufer_source_counting_2018,hsu2023learning}.
While effective, these aforementioned methods
and also recent \gls{DNN} models that jointly perform source counting and separation~\cite{wang2021count,saijo2023single}
typically estimate the source count over long signal segments ($1-\SI{20}{\second}$),
which is too slow for low-latency speech communication applications.
Other methods tackling online speaker counting~\cite{grumiaux2021high,yousefi2021real,cornell2022overlapped}
% operate with latencies of $\geq\SI{200}{\milli\second}$ or rely on Ambisonics features~\cite{grumiaux2021high},
operate with latencies of at least $\SI{200}{\milli\second}$ or rely on Ambisonics features~\cite{grumiaux2021high},
which is not directly applicable to binaural hearing aids.
As a more suitable baseline for online processing, we will consider
the framewise source counting method proposed in~\cite{Gode2025EBOP},
using the \gls{GMSC}~\cite{gmsc_ramirez_2008} on spatially whitened signals.
However, this method is limited to detecting only source activations
and relies on manually-tuned thresholds.

\begin{figure}
  \centering
  \includegraphics[width=\columnwidth]{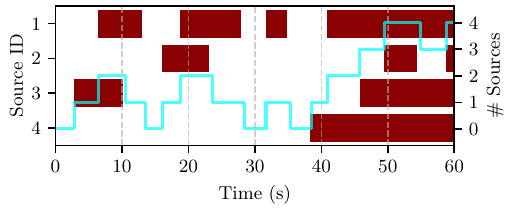}
  \vspace{-18pt}
  \caption{Exemplary source activity timeline with 4 sources,
    showing activity intervals and the total number of active sources.}
  \label{fig:source_activity_timeline}
\end{figure}

Inspired by the \gls{GMSC}-based method in~\cite{Gode2025EBOP},
in this paper we propose a more robust \gls{DNN}-based source counting method.
First, we introduce a novel set of features designed to detect source deactivations,
based on a time-reversed whitening approach.
% By combining these new deactivation features with the established activation features,
% we create a comprehensive feature representation
% that is sensitive to both increases and decreases in the source count.
Second, to overcome the limitations of manually tuned thresholds,
we train compact neural networks to provide a framewise estimate of the number of active sources
based on the spatial coherence features in the whitened domain.
We consider two causal architectures, a \gls{TCN} and a \gls{GRU}-based \acrlong{RNN},
The proposed systems are evaluated for a binaural hearing aid setup
on recorded noisy and reverberant acoustic scenarios
with a time-varying number of speakers
(see exemplary scenario in~\Cref{fig:source_activity_timeline}).
Results show that the proposed \gls{DNN}-based source count estimators significantly outperform
the conventional threshold-based method in terms of accuracy and \acrlong{MAE},
and especially benefit from
incorporating the proposed deactivation features.
The \gls{GRU}-based source count estimator consistently outperforms the \gls{TCN}-based source count estimator,
achieving a frame-wise accuracy of \SI{91.9}{\percent}.

\vspace{\abovesectionvspace}
\section{Signal Model}
\label{sec:signal_model}
\vspace{\belowsectionvspace}

We consider an acoustic scenario with a time-varying number
of spatially-stationary sound sources in a noisy and reverberant environment,
recorded by an array of $M$ microphones.
In the \gls{STFT} domain, the $m$-th microphone signal is denoted by $y_{m,t,f}$,
with time frame $t$ and frequency $f$.
The microphone signals are stacked in the vector
\begin{equation}
  \mathbf{y}_{t,f} =
  \begin{bmatrix}
    y_{1,t,f} & \cdots & y_{M,t,f}
  \end{bmatrix}^{\mathrm{T}} \in \mathbb{C}^{M},
  \label{eq:sig_vec_def}
\end{equation}
where $\{\cdot\}^{\mathrm{T}}$ denotes the transpose.
For notational simplicity, the frequency index $f$ will be omitted, except when it is explicitly required. % TODO: maybe remove this sentence and put the frequency index everywhere
The microphone signal vector $\mathbf{y}_{t}$
is modeled as the sum of all active source contributions and additive noise, i.e.,
\begin{equation}
  \mathbf{y}_{t} =
  % \underbrace{
  \sum_{k=1}^{K_{\text{max}}} \mathcal{I}_{k,t} \mathbf{x}_{k,t}
  % }_{\mathbf{x}_{t}}
  + \mathbf{n}_{t},
  \label{eq:signal_model}
\end{equation}
where $K_{\text{max}}$ denotes the total number of potential sources,
$\mathbf{x}_{k,t} \in \mathbb{C}^{M}$ denotes the $k$-th source vector,
and $\mathbf{n}_{t} \in \mathbb{C}^{M}$ denotes the noise vector.
The frequency-independent binary indicator $\mathcal{I}_{k,t} \in \{0,1\}$
denotes the (broadband) activity of the $k$-th source at time frame $t$,
modeling the time-varying number of active sources.
The method considered in this paper aims at estimating the number of active sources
\begin{equation}
  K_{t} = \sum_{k=1}^{K_{\text{max}}} \mathcal{I}_{k,t},
  \label{eq:active_source_count}
\end{equation}
per frame using only $\mathbf{y}_{t}$ from current and past frames.

Assuming that the length of the \gls{STFT} analysis window is sufficiently large compared to the length of the \acrlong{RIR},
% the convolution of the time domain source signal with the \gls{RIR}
% can be approximated by a multiplication in the \gls{STFT} domain~\cite{avargel_multiplicative_2007}.
% Consequently,
each source component $\mathbf{x}_{k,t}$ can be written as~\cite{avargel_multiplicative_2007}
\begin{equation}
  \mathbf{x}_{k,t} = \mathbf{a}_{k} s_{k,t},
  \label{eq:ATF_source_model}
\end{equation}
where $s_{k,t}$ denotes the \gls{STFT} coefficient of the $k$-th source signal
and $\mathbf{a}_{k} \in \mathbb{C}^{M}$ denotes the time-invariant \gls{ATF} vector from the $k$-th source to the microphone array.
The \gls{ATF} vectors of all sources are assumed to be linearly independent.
Assuming statistical independence between all sources and the noise,
the covariance matrix of the microphone signals
$\mathbf{R}_{y,t} = \mathbb{E}\{\mathbf{y}_{t} \mathbf{y}_{t}^{\mathrm{H}}\}$ can be expressed as
\begin{equation}
  \mathbf{R}_{y,t} =
  % \underbrace{
  \sum_{k=1}^{K_{\text{max}}} \mathcal{I}_{k,t} \phi_{k,t} \mathbf{a}_{k} \mathbf{a}_{k}^{\mathrm{H}}
  % }_{\mathbf{R}_{x,t}}
  + \mathbf{R}_{n}
  \quad\in\mathbb{C}^{M\times M},
  \label{eq:cov_model}
\end{equation}
where $\{\cdot\}^{\mathrm{H}}$ denotes the conjugate transpose,
$\phi_{k,t} = \mathbb{E}\{|s_{k,t}|^2\}$ denotes the \gls{PSD} of the $k$-th source,
% $\mathbf{R}_{x,t} = \mathbb{E}\{\mathbf{x}_{t} \mathbf{x}_{t}^{\mathrm{H}}\}$
% denotes the covariance matrix of the source signals,
and $\mathbf{R}_{n} = \mathbb{E}\{\mathbf{n}_{t} \mathbf{n}_{t}^{\mathrm{H}}\}$ denotes the noise covariance matrix,
assumed to be time-invariant.
% In practice, the covariance matrix $\mathbf{R}_{y,t}$ is estimated from the observed microphone signals.
% A common approach is to recursively update an estimate $\widehat{\mathbf{R}}_{y,t}$ at each time frame $t$ using an exponential sliding window, i.e.,
% \begin{equation}
%   \widehat{\mathbf{R}}_{y,t} = \alpha \widehat{\mathbf{R}}_{y,t-1} + (1-\alpha)\mathbf{y}_{t}\mathbf{y}_{t}^{\mathrm{H}},
%   \label{eq:cov_est}
% \end{equation}
% where the forgetting factor $\alpha = e^{-t_{\mathrm{fs}}/t_{\alpha}}$ depends on the \gls{STFT} frame shift $t_{\mathrm{fs}}$ and a smoothing time constant $t_{\alpha}$.

Considering the activation or deactivation of a single source $k'$ between two consecutive time frames
$t-1$ and $t$, the number of active sources in~\Cref{eq:active_source_count} changes by one, i.e., $K_{t} = K_{t-1} \pm 1$.
Assuming the \glspl{PSD} of the other active sources do not change significantly between consecutive time frames,
the matrix $\mathbf{R}_{y,t}$ in~\Cref{eq:cov_model} can be expressed as
\begin{equation}
  \boxed{
  \mathbf{R}_{y,t} \approx \mathbf{R}_{y,t-1} \pm \phi_{k'} \mathbf{a}_{k'} \mathbf{a}_{k'}^{\mathrm{H}}
  }
  \label{eq:cov_change}
\end{equation}
where $\phi_{k'}$ denotes the \gls{PSD} of the $k'$-th source (in frame $t$ for activation and frame $t-1$ for deactivation).
It can be observed in \Cref{eq:cov_change} that a source activation corresponds to the addition of a rank-1 matrix,
whereas a source deactivation corresponds to the subtraction of a rank-1 matrix.
The core idea of the spatial-coherence based source counting methods,
in this paper is to utilize these rank-1 updates of the noisy covariance matrix
to track the number of active sources $K_t$.

\vspace{\abovesectionvspace}
\section{Conventional Source Activation Detection}
\label{sec:conventional}
\vspace{\belowsectionvspace}

The conventional method for source activation detection~\cite{Gode2025EBOP}
directly exploits the rank-1 update property in \Cref{eq:cov_change}.
The core idea is to reformulate the problem of estimating the number of sources per time frame
into the problem of detecting the activation of a single coherent source in spatially white noise.
This is achieved by a whitening process
(taking into account already active sources and noise),
so that the spatial coherence in this whitened domain,
is expected to increase significantly only when a new source activates.

\begin{figure*}
  \centering
  \includegraphics[width=\linewidth]{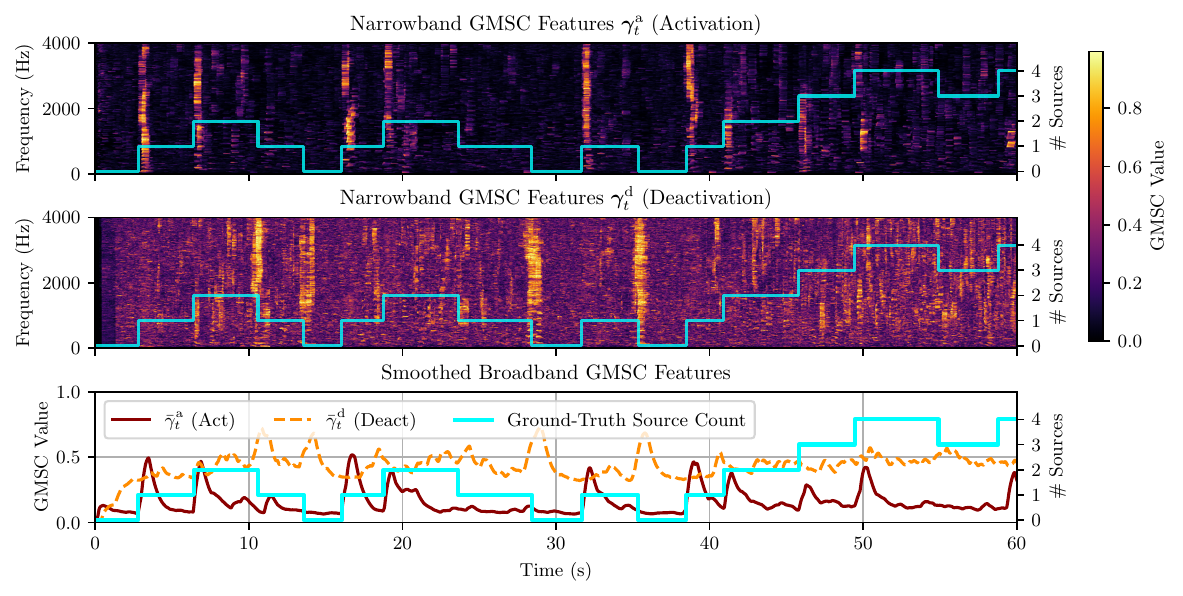}
  \vspace{-24pt}
  \caption{Illustration of the GMSC-based features (\gls{SNR}$=\SI{9.5}{\decibel}$). Top: GMSC features for activation.
    Middle: GMSC features for deactivation. Bottom: Corresponding broadband features after recursive smoothing.}
  \vspace{-12pt}
  \label{fig:feature_figure}
\end{figure*}

\vspace{\abovesubsectionvspace}
\subsection{Whitened GMSC Features for Source Activation Detection}
\label{ssec:wgmsc4activation}
\vspace{\belowsubsectionvspace}

To detect the addition of the rank-1 matrix corresponding to a new source activation,
the conventional method~\cite{Gode2025EBOP} whitens the covariance matrix $\mathbf{R}_{y,t}$
at time frame $t$ with a reference covariance matrix denoted as $\mathbf{R}_{v,t}$.
Using a matrix square-root decomposition of $\mathbf{R}_{v,t}$, i.e.,
$\mathbf{R}_{v,t} = \mathbf{R}_{v,t}^{\nicefrac{\mathrm{H}}{2}}\mathbf{R}_{v,t}^{\nicefrac{1}{2}}$ (e.g., Cholesky decomposition),
the whitened covariance matrix $\mathbf{R}^{\text{a}}_{w,t}$ is computed as
\begin{equation}
  \boxed{
  \mathbf{R}^{\text{a}}_{w,t} =
  \mathbf{R}_{v,t}^{-\nicefrac{\mathrm{H}}{2}}
  \mathbf{R}_{y,t}
  \mathbf{R}_{v,t}^{-\nicefrac{1}{2}}
  }
  \label{eq:whitening_for_sad}
\end{equation}
The reference covariance matrix is used as an estimate
of the covariance matrix before the activation,
containing the $K_{t}-1$ already active sources and the noise.
In~\cite{Gode2025EBOP} it was proposed to use the covariance matrix $\mathbf{R}_{y,t}$
from $t_v$ frames in the past as the reference covariance matrix, i.e.,
\begin{equation}
  \mathbf{R}_{v,t} = \mathbf{R}_{y,t-t_v},
  \label{eq:undesired_cov_est}
\end{equation}
% where $t_v = \nicefrac{t_{\text{r}}}{t_{\text{fs}}}$ and $t_{\text{r}}$ denotes the assumed minimum time between two source activations.
where $t_v$ corresponds to the assumed minimum time between two source activations.
As long as no new source activates,
$\mathbf{R}_{y,t} \approx \mathbf{R}_{v,t}$,
such that the whitened covariance matrix $\mathbf{R}^{\text{a}}_{w,t}$ is approximately equal to the identity matrix,
resulting in low spatial coherence.
When a new source becomes active, $\mathbf{R}^{\text{a}}_{w,t}$ contains an additional rank-1 component,
leading to a sharp increase in spatial coherence.
To quantify this, it was proposed in~\cite{Gode2025EBOP} to use the \gls{GMSC}~\cite{gmsc_ramirez_2008},
which extends the concept of spatial coherence between two microphones to multiple microphones.
The \gls{GMSC} is defined as
\begin{equation}
  \gamma_{t}^{\text{a}} = \frac{\lambda_{\text{max}}\left\{\bm{\Gamma}_{t}^{\text{a}}\right\}-1}{M-1},
  \label{eq:gmsc}
\end{equation}
yielding a value between $0$ and $1$, where $\lambda_{\text{max}}\{\cdot\}$ denotes the principal eigenvalue
and $\bm{\Gamma}_{t}^{\text{a}}$ denotes the coherence matrix of the whitened signals, i.e.,
\begin{equation}
  \bm{\Gamma}_{t}^{\text{a}} = \mathbf{D}_{w,t}^{-\nicefrac{1}{2}} \mathbf{R}^{\text{a}}_{w,t} \mathbf{D}_{w,t}^{-\nicefrac{1}{2}}.
  \label{eq:coherence_matrix}
\end{equation}
where $\mathbf{D}_{w,t}$ denotes a diagonal matrix containing the diagonal entries of $\mathbf{R}^{\text{a}}_{w,t}$.
Although presented without a frequency index for notational simplicity,
the whitening in \Cref{eq:whitening_for_sad} and the \gls{GMSC} calculation in \Cref{eq:coherence_matrix,eq:gmsc}
are performed for each frequency $f$.
The resulting frequency-dependent \gls{GMSC} values $\gamma_{t,f}^{\text{a}}$ are then stacked into the \gls{GMSC} feature vector
\begin{equation}
  \bm{\gamma}^{\text{a}}_{t} = 
  \begin{bmatrix}
    \gamma_{t,1}^{\text{a}} & \cdots & \gamma_{t,F}^{\text{a}}
  \end{bmatrix}^{\mathrm{T}} \in \mathbb{R}^{F},
  \label{eq:gmsc_feature_vector}
\end{equation}
where $F$ denotes the number of frequency bins.
This feature vector, whose dimension does not depend on the number of microphones $M$,
serves as the input for the subsequent detection stages.
% This process is illustrated by the solid path in the block diagram in \Cref{fig:block_diagram}.
For an exemplary acoustic scenario, the top subplot of \Cref{fig:feature_figure}
illustrates the \gls{GMSC}-based activation features.
It can be observed that high \gls{GMSC} values occur at time frames when new sources activate.

In practice, the covariance matrix $\mathbf{R}_{y,t}$ is estimated from the observed microphone signals
by recursive smoothing, i.e.,
\begin{equation}
  \widehat{\mathbf{R}}_{y,t} = \alpha \widehat{\mathbf{R}}_{y,t-1} + (1-\alpha)\mathbf{y}_{t}\mathbf{y}_{t}^{\mathrm{H}},
  \label{eq:cov_est}
\end{equation}
where the forgetting factor $\alpha = e^{-t_{\mathrm{fs}}/t_{\alpha}}$
depends on the \gls{STFT} frame shift $t_{\mathrm{fs}}$ and a smoothing time constant $t_{\alpha}$.

\vspace{\abovesubsectionvspace}
\subsection{Threshold-Based Detection}
\label{ssec:threshold_detection}
\vspace{\belowsubsectionvspace}

For the conventional threshold-based source activation detection~\cite{Gode2025EBOP},
the feature vector $\bm{\gamma}^{\text{a}}_{t}$ is first condensed
into a single broadband value $\tilde{\gamma}^{\text{a}}_{t}$
by computing a weighted average, i.e.,
\begin{equation}
  \tilde{\gamma}^{\text{a}}_{t} = \mathbf{w}_{t}^{\mathrm{T}} \bm{\gamma}^{\text{a}}_{t},
  \label{eq:gmsc_broadband}
\end{equation}
where the weight vector
$\mathbf{w}_{t} =
  \begin{bmatrix}
    w_{t,1} & \cdots & w_{t,F}
  \end{bmatrix}^{\mathrm{T}} \in \mathbb{R}^{F}$
contains the normalized power of the whitened signal in each frequency bin,
i.e., $w_{t,f} = \mathrm{tr}\{\mathbf{R}_{w,t,f}^{\text{a}}\} / \sum_{f=1}^{F} \mathrm{tr}\{\mathbf{R}_{w,t,f}^{\text{a}}\}$,
where $\mathrm{tr}\{\cdot\}$ denotes the trace.
The broadband \gls{GMSC} $\tilde{\gamma}^{\text{a}}_{t}$ serves as a direct indicator of a new source activation.
To improve robustness, $\tilde{\gamma}^{\text{a}}_{t}$ is recursively smoothed, i.e.,
\begin{equation}
  \bar{\gamma}^{\text{a}}_{t} = \beta \bar{\gamma}_{t-1} + (1-\beta) \tilde{\gamma}^{\text{a}}_{t},
  \label{eq:gmsc_smoothing}
\end{equation}
where the forgetting factor $\beta = e^{-t_{\mathrm{fs}}/t_{\gamma}}$ depends on
the \gls{STFT} frame shift $t_{\mathrm{fs}}$ and a smoothing time constant $t_{\gamma}$.
An example of the smoothed broadband \gls{GMSC} $\bar{\gamma}^{\text{a}}_{t}$ is shown in the bottom subplot of \Cref{fig:feature_figure} (red curve).
A source activation is then detected
if the smoothed feature $\bar{\gamma}^{\text{a}}_{t}$ exceeds a predefined threshold $\gamma^{\text{a}}_{\tau}$.
Upon detection, the source count $K_{t}$ is incremented by one.
It should be noted that determining a single threshold
that performs robustly across diverse acoustic scenarios
(e.g., varying SNRs or source positions) is non-trivial.

\vspace{\abovesectionvspace}
\section{Source Deactivation Detection Features}
% \section{Proposed Whitened GMSC Features for Source Deactivation Detection} TOOO: if space permits this section title could be used
\label{sec:gmsc4deactivation}
\vspace{\belowsectionvspace}

A disadvantage of the conventional source activity estimation method~\cite{Gode2025EBOP}
is that it can only detect source activations,
but no source deactivations,
which is obviously unrealistic.
To extend the conventional method to also detect source deactivations,
we propose a deactivation feature based on interpreting a deactivation as a time-reversed activation.
Following this intuition, a similar whitening-based approach as in~\Cref{ssec:wgmsc4activation}
can be used by swapping the roles of the current and reference covariance matrices.
Instead of whitening the covariance matrix $\mathbf{R}_{y,t}$ at frame $t$
with the covariance matrix $\mathbf{R}_{y,t-t_{v}}$ from the past,
we whiten the covariance matrix $\mathbf{R}_{y,t-t_{v}}$ from the past
with the current covariance matrix $\mathbf{R}_{y,t}$, i.e.,
% Let us consider a deactivation at time frame $t-t_{v}$.
% We now use the current covariance matrix,
% denoted as $\mathbf{R}'_{y,t}$ to distinguish it from the activation case,
% as the reference for whitening,
% assuming it does not include any information about the deactivated source.
% The matrix to be whitened is the one from $t_{v}$ frames in the past,
% i.e., $\mathbf{R}'_{y,t-t_{v}}$, which contains the $K_{t}$ active sources
% including the one that deactivated at time $t-t_{v}$.
% The whitened covariance matrix $\mathbf{R}^{\text{d}}_{w,t}$ is then computed similarly to~\Cref{eq:whitening_for_sad} as
\begin{equation}
  \boxed{
  \mathbf{R}^{\text{d}}_{w,t} =
  \mathbf{R}_{y,t}^{-\nicefrac{\mathrm{H}}{2}}
  \mathbf{R}_{y,t-t_v}
  \mathbf{R}_{y,t}^{-\nicefrac{1}{2}}
  }
  \label{eq:whitening_for_deact}
\end{equation}
As long as no source deactivates, $\mathbf{R}_{y,t} \approx \mathbf{R}_{y,t-t_v}$,
such that the whitened covariance matrix $\mathbf{R}^{\text{d}}_{w,t}$
is approximately equal to the identity matrix,
resulting in a low \gls{GMSC} value.
If a source deactivates at time $t-t_v$,
then %$\mathbf{R}'_{y,t-t_v} \approx \mathbf{R}'_{y,t} + \phi_{k'} \mathbf{a}_{k'} \mathbf{a}_{k'}^{\mathrm{H}}$.
%In this case,
$\mathbf{R}^{\text{d}}_{w,t}$ becomes approximately equal to an identity matrix plus a rank-1 component,
leading to a high \gls{GMSC} value.
This effectively turns deactivation detection into an activation detection problem,
with an inherent processing delay of $t_v$ frames.

Due to this inherent processing delay,
we decided not to use recursive smoothing for estimating the covariance matrix
as in~\Cref{eq:cov_est}, but to use a more instantaneous moving-average estimate, i.e.,
\begin{equation}
  \widehat{\mathbf{R}}_{y,t} = \frac{1}{L} \sum_{l=0}^{L-1} \mathbf{y}_{t-l}\mathbf{y}_{t-l}^{\mathrm{H}}.
  \label{eq:cov_est_deact}
\end{equation}
where the sliding window length $L$ is chosen to be smaller than $t_v$.

% It should be noted that the recursive estimation of the microphone covariance matrix from \Cref{eq:cov_est}
% cannot be used for this time-reversed method.
% Its long temporal decay would violate the core assumption
% that the current covariance matrix is free from components corresponding to the deactivated source.
% To ensure a clean separation, we instead estimate the covariance matrices for the deactivation features
% using a moving average over a finite-length sliding window of $L$ frames:
% \begin{equation}
%   \widehat{\mathbf{R}}'_{y,t} = \frac{1}{L} \sum_{l=0}^{L-1} \mathbf{y}_{t-l}\mathbf{y}_{t-l}^{\mathrm{H}}.
%   \label{eq:cov_est_ma}
% \end{equation}
% By choosing the delay $t_v \ge L$, the estimation windows for the current and past covariance matrices
% are guaranteed to be non-overlapping, thus strictly satisfying the assumption.

From the whitened matrix $\mathbf{R}^{\text{d}}_{w,t}$, the feature vector $\bm{\gamma}^{\text{d}}_{t}$ for deactivation
is computed similarly to~\Cref{eq:gmsc}~-~\Cref{eq:gmsc_feature_vector}.
For the deactivation detection, a similar threshold-based method as described in \Cref{ssec:threshold_detection}
can be applied to the averaged and smoothed broadband value of this feature vector $\bar{\gamma}^{\text{d}}_{t}$ (computed similarly to \Cref{eq:gmsc_broadband} and \Cref{eq:gmsc_smoothing}),
where a detection now leads to decrementing the source count $K_t$ by one.
% This process is illustrated by the dashed path in the block diagram in \Cref{fig:block_diagram}.
An example of the proposed \gls{GMSC} deactivation feature vector $\bm{\gamma}^{\text{d}}_{t}$
and the smoothed broadband \gls{GMSC} $\bar{\gamma}^{\text{d}}_{t}$
is illustrated in the middle and bottom subplots (orange curve) of \Cref{fig:feature_figure}.

\vspace{\abovesectionvspace}
\section{Proposed DNN-Based Source Count Estimation}
\label{sec:proposed_dnn}
\vspace{\belowsectionvspace}

To overcome the limitations of the threshold-based method,
which requires careful manual tuning and may not generalize well,
we propose to train a \gls{DNN} to estimate
the number of active sources $K_t$ at each time frame from spatial coherence features.
As input, we use the \gls{GMSC}-based feature vectors
$\bm{\gamma}^{\text{a}}_{t}$ and $\bm{\gamma}^{\text{d}}_{t}$ for activation and deactivation,
which are concatenated to form a combined feature vector
$\bm{\zeta}_{t} \in \mathbb{R}^{2F}$.
% \begin{equation}
%   \bm{\zeta}_{t} = [\bm{\gamma}_{t}^{\mathrm{T}}, \bm{\gamma}'^{\mathrm{T}}_{t}]^{\mathrm{T}} \in \mathbb{R}^{2F}.
%   \label{eq:combined_feature_vector}
% \end{equation}
We investigate two causal network architectures:
a \gls{TCN}~\cite{luo2019conv}, which uses dilated convolutions to create a large receptive field,
and a \gls{GRU}-based \gls{RNN}~\cite{cho2014properties}, which uses its internal state to capture temporal dependencies.
For both architectures, a final softmax layer outputs probabilities $p_{t,k}$
for each possible source count $k \in \{0, \dots, K_{\text{max}}\}$,
and the final estimate is taken as the most likely class, i.e.,
$
  %\begin{equation}
  \widehat{K}_t = \underset{k}{\arg\max} \;\; p_{t,k}.
  %\label{eq:dnn_output}
  %\end{equation}
$

\vspace{\abovesectionvspace}
\section{Evaluation}
\label{sec:evaluation}
\vspace{\belowsectionvspace}

In this section, the performance of the conventional threshold-based method
and the proposed \gls{DNN}-based source count estimators is
compared for several acoustic scenarios with up to $4$ speakers and background noise.
The experimental setup, including dataset generation, algorithmic implementation, and training procedure, is detailed below.

\vspace{\abovesubsectionvspace}
\subsection{Dataset}
\label{ssec:dataset}
\vspace{\belowsubsectionvspace}

\begin{figure}
  \centering
  \includegraphics[width=\columnwidth]{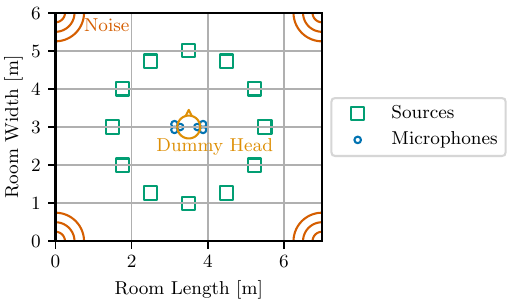}
  \vspace{-18pt}
  \caption{Acoustic setup of the BRUDEX database~\cite{fejgin2023brudex}.}
  \label{fig:dataset_examples}
\end{figure}

Acoustic scenarios were generated by convolving clean speech from Librispeech~\cite{panayotov2015librispeech}
with binaural \glspl{RIR} from the BRUDEX database~\cite{fejgin2023brudex}
for the low-reverberation condition ($\text{T}_{60} \approx \SI{310}{\milli\second}$)
at a sampling frequency of $\SI{8}{\kilo\hertz}$.
As depicted in~\Cref{fig:dataset_examples},
the \glspl{RIR} cover 12 source positions on a circle of radius $\SI{1}{\meter}$ around a dummy head with an angular spacing of $\SI{30}{\degree}$.
The dummy head was equipped with a \gls{BTE} hearing aid on each side,
with each \gls{BTE} having a front and a rear microphone spaced $\SI{1.5}{\centi\meter}$ apart.
Additionally, an in-ear microphone was present on each side of the head.
This setup provides various microphone array configurations using 2 to 6 of these microphones.
Quasi-diffuse babble, cafeteria or white noise from BRUDEX was added at a random \gls{SNR} between $\SI{5}{\decibel}$ and $\SI{15}{\decibel}$.
To create dynamic scenes, the number of simultaneously active sources varies up to $K_{\text{max}}=4$,
with activation/deactivation events occurring at random intervals between $\SI{2}{\second}$ and $\SI{5}{\second}$.
Based on this, two distinct datasets were generated:
Dataset A ($\SI{20}{\second}$ scenarios) contains only source activations,
while Dataset B ($\SI{60}{\second}$ scenarios) contains both activations and deactivations.
Each dataset is split into $6000$ training, $600$ validation, and $600$ test scenarios,
where for each scenario the source positions, clean speech signals,
microphone array configuration, SNR, noise type, and activity pattern were chosen randomly.

\vspace{\abovesubsectionvspace}
\subsection{Algorithmic Implementation and Training}
\label{ssec:implementation}
\vspace{\belowsubsectionvspace}

The microphone signals are processed using an \gls{STFT} framework
with a frame length of $800$ samples (corresponding to $\SI{100}{\milli\second}$),
resulting in $F=401$,
a frame shift of $\SI{25}{\percent}$, i.e., $t_{\text{fs}}=\SI{25}{\milli\second}$,
and a square-root Hann analysis window.
The GMSC-based activation features are computed with a recursive smoothing of $t_{\alpha} = \SI{500}{\milli\second}$ in~\Cref{eq:cov_est},
while the deactivation features are computed with a sliding window of $L=14$
(corresponding to $\SI{350}{\milli\second}$) in~\Cref{eq:cov_est_deact}.
For both features, the delay for the reference covariance matrix in~\Cref{eq:undesired_cov_est,eq:whitening_for_deact}
is equal to $t_{v} = 20$ frames (corresponding to $\SI{500}{\milli\second}$).

We compare three source count estimators using either only activation features for Dataset A and Dataset B,
or using concatenated activation and deactivation features for Dataset B:
\begin{itemize}[leftmargin=*,topsep=1pt,itemsep=1pt,partopsep=1pt,parsep=1pt]
  % \vspace{-3pt}
  \item The conventional \textbf{threshold-based} method using broadband feature smoothing of $t_{\gamma}=\SI{500}{\milli\second}$ in~\Cref{eq:gmsc_smoothing}
        and fixed thresholds of $\gamma^{\text{a}}_{\tau}=0.24$ (activation) and $\gamma^{\text{d}}_{\tau}=0.62$ (deactivation),
        which were determined using a grid search on the test set of Dataset A and Dataset B, respectively.
        % \vspace{-3pt}
  \item An \textbf{\gls{TCN}-based} estimator using 3 stacks of 3 layers with a kernel size of 3,
        a bottleneck dimension of 128, and a hidden dimension of 256, resulting in a receptive field of 43 frames ($=\SI{1.075}{\second}$).% and about 1 million parameters.
        % \vspace{-3pt}
  \item An \textbf{\gls{RNN}-based} estimator using a 3-layer \gls{GRU} with a hidden size of half the input dimension.
        % \vspace{-3pt}
\end{itemize}
The \gls{DNN} models were trained for 100 epochs using the AdamW optimizer
with a learning rate of $10^{-4}$ and a batch size of 30 to minimize the cross-entropy loss
between the estimated and the ground-truth source count.
This ground-truth, $K_t$ was computed using~\Cref{eq:active_source_count},
where the ground-truth activity indicator $\mathcal{I}_{k,t}$
was determined by applying a power-based voice activity detector
to the source components $\mathbf{x}_{k,t}\;\forall k\in\{1,\ldots,K_{\text{max}}\}$.

\vspace{\abovesubsectionvspace}
\subsection{Results}
\label{ssec:results}
\vspace{\belowsubsectionvspace}

\begin{table}[t!]
  \centering
  \vspace{\tablecapvspace}
  \caption{Accuracy and \gls{MAE} results on Dataset A and Dataset B for different feature sets ($\bm{\gamma}^{\text{a}}_{t}$ or $\bm{\zeta}_t$).}
  \label{tab:results}
  \begin{tabular}{llcc}
    \toprule
    \textbf{Features}                               & \textbf{Detector Method} & \textbf{Accuracy [\si{\percent}]} & \textbf{MAE}  \\
    \midrule
    \multicolumn{4}{l}{$\quad\quad\quad$\textbf{Dataset A} (Activations only)}                                                     \\
    \midrule
    \textit{Features: $\bm{\gamma}^{\text{a}}_{t}$} & Threshold-based          & 83.1                              & 0.18          \\
                                                    & TCN-based                & 94.2                              & 0.06          \\
                                                    & RNN-based                & \textbf{95.6}                     & \textbf{0.04} \\
    \midrule
    \multicolumn{4}{l}{$\quad\quad\quad$\textbf{Dataset B} (Activations \& Deactivations)}                                         \\
    \midrule
    \textit{Features: $\bm{\gamma}^{\text{a}}_{t}$} %& Threshold-based & 78                                & 0.35           \\
                                                    & TCN-based                & 83.8                              & 0.17          \\
                                                    & RNN-based                & \textbf{89.0}                     & \textbf{0.12} \\
    \midrule
    \textit{Features: $\bm{\zeta}_t$}               & Threshold-based          & 33.6                              & 1.21          \\
                                                    & TCN-based                & 86.4                              & 0.14          \\
                                                    & RNN-based                & \textbf{91.9}                     & \textbf{0.09} \\
    \bottomrule
  \end{tabular}
  \vspace{-3pt}
\end{table}
Performance is evaluated on the test sets using two framewise metrics:
the classification accuracy (percentage of frames where $\widehat{K}_t = K_t$)
and the \acrfull{MAE}, i.e.,
$
  \text{MAE} = \nicefrac{1}{T} \sum_{t=1}^{T} | \widehat{K}_t - K_t |
$,
where $T$ is the total number of frames in a test set. The results are presented in \Cref{tab:results}.
On the activation-only Dataset A, both \gls{DNN}-based methods significantly outperform the threshold-based method,
with the \gls{RNN} achieving the best accuracy of \SI{95.6}{\percent}.
On the more challenging Dataset B, which includes both activations and deactivations,
the performance of the \gls{DNN} models using only activation features $\bm{\gamma}^{\text{a}}_{t}$ is,
as expected, lower than on Dataset A. The benefit of the proposed deactivation features $\bm{\gamma}^{\text{d}}_{t}$ is clearly demonstrated:
while the threshold-based method fails completely on Dataset B (accuracy of \SI{33.6}{\percent}),
both \gls{DNN}-based methods show a notable performance improvement compared to only using the activation features $\bm{\gamma}_{t}^{\text{a}}$.
The \gls{RNN}-based estimator consistently performs best,
reaching an accuracy of $\SI{91.9}{\percent}$ when using both activation and deactivation features.
Finally, the proposed method is real-time feasible, with an average real-time factor of $0.007$ (dominated by feature extraction) measured on an NVIDIA RTX A5000 GPU.
\vspace{-3pt}

\vspace{\abovesectionvspace}
\section{Conclusion}
\label{sec:conclusion}
\vspace{\belowsectionvspace}
\vspace{-3pt}

This paper extends a spatial coherence-based method for online source activation detection.
We propose \gls{GMSC}-based features for detecting source deactivations
using a time-reversed whitening approach, complementing existing activation features.
Furthermore, we propose to replace the conventional threshold-based source count estimator
with more robust \gls{DNN}-based classifiers (\gls{TCN} and \gls{GRU})
that directly estimate the source count from the combined feature set.
The evaluation on real-world acoustic scenarios shows that the \gls{DNN}-based methods
significantly outperform the threshold-based method.
Crucially, incorporating the proposed deactivation features substantially improves performance
in scenarios involving both activations and deactivations.
The \gls{RNN}-based source count estimator proves most effective,
achieving a frame-wise accuracy of \SI{91.9}{\percent}.

% Below is an example of how to insert images. Delete the ``\vspace'' line,
% uncomment the preceding line ``\centerline...'' and replace ``imageX.ps''
% with a suitable PostScript file name.
% -------------------------------------------------------------------------
% \begin{figure}[htb]

%   \begin{minipage}[b]{1.0\linewidth}
%     \centering
%     \centerline{\includegraphics[width=8.5cm]{ICASSP2026_Paper_Templates/image1}}
%     %  \vspace{2.0cm}
%     \centerline{(a) Result 1}\medskip
%   \end{minipage}
%   %
%   \begin{minipage}[b]{.48\linewidth}
%     \centering
%     \centerline{\includegraphics[width=4.0cm]{ICASSP2026_Paper_Templates/image3}}
%     %  \vspace{1.5cm}
%     \centerline{(b) Results 3}\medskip
%   \end{minipage}
%   \hfill
%   \begin{minipage}[b]{0.48\linewidth}
%     \centering
%     \centerline{\includegraphics[width=4.0cm]{ICASSP2026_Paper_Templates/image4}}
%     %  \vspace{1.5cm}
%     \centerline{(c) Result 4}\medskip
%   \end{minipage}
%   %
%   \caption{Example of placing a figure with experimental results.}
%   \label{fig:res}
%   %
% \end{figure}

% \vfill
% \pagebreak

\bibliographystyle{IEEEbib}
\bibliography{refs}

\end{document}